\documentclass[
 reprint,
showpacs,amsmath,amssymb,
 aps, pra,]{revtex4-1}

\usepackage{graphicx}
\usepackage{dcolumn}
\usepackage{bm}
\usepackage{bigstrut}

\begin{document}

\preprint{APS/123-QED}

\title{Sensitivity of rotational transitions in CH and CD to a possible variation of fundamental constants}

\author{Adrian J. de Nijs}
\author{Wim Ubachs}
\author{Hendrick L. Bethlem}
\affiliation{LaserLaB, Department of Physics and Astronomy, VU University, De Boelelaan 1081, 1081 HV Amsterdam, The Netherlands}

\date{\today}

\begin{abstract}
The sensitivity of rotational transitions in CH and CD to a possible variation of fundamental constants has been investigated. Largely enhanced sensitivity coefficients are found for specific transitions which are due to accidental degeneracies between the different fine-structure manifolds. These degeneracies occur when the spin-orbit coupling constant is close to four times the rotational constant. CH and particularly CD match this condition closely. Unfortunately, an analysis of the transition strengths shows that the same condition that leads to an enhanced sensitivity suppresses the transition strength, making these transitions too weak to be of relevance for testing the variation of fundamental constants over cosmological time scales. We propose a test in CH based on the comparison between the rotational transitions between the $e$ and $f$ components of the $\Omega'=1/2,J=1/2$ and $\Omega'=3/2,J=3/2$ levels at $532$ and $536$~GHz and other rotational or Lambda-doublet transitions in CH involving the same absorbing ground levels. Such a test, to be performed by radioastronomy of highly redshifted objects, is robust against systematic effects.
\end{abstract}

\pacs{33.20.-t,06.20.Jr}

\maketitle

\section{Introduction}

A possible variation of the fundamental constants can be detected by comparing transitions between levels in atoms and molecules that have a different functional dependence on these constants. The limit that can be derived from such a test is proportional to the relative accuracy of the experiment and inversely proportional to both the time interval covered by the experiment and the sensitivity of the transition to a possible variation. The duration of tests that are conducted in the laboratory is typically limited to a few years, but these tests have the advantage that one can choose transitions in atoms or molecules that are either very sensitive to a variation, transitions that can be measured to an extremely high precision, or both. Tests over cosmological time scales, on the other hand, typically span 10$^{9}$ years, but have the disadvantage that only a limited number of molecular transitions are observed at high redshift, and the accuracy of the observed lines is relatively low.

Up to very recently, tests of the time-variation of the proton-to-electron mass ratio, $\mu=m_p/m_e$, over cosmological time scales were based exclusively on molecular hydrogen, the most abundant molecule in the universe and observed in a number of high redshift objects. The transitions in molecular hydrogen correspond to transitions between different electronic states and exhibit sensitivity coefficients, $K_{\mu}$, ranging from $-0.05$ to $+0.01$~\cite{ubachs:jms,weerdenburg:prl}. Recently, the observations of the inversion transition in ammonia ($K_{\mu}=-4.2$)~\cite{vanVeldhoven,Flambaum:PRL2007,Murphy:Science} and torsion-rotation transitions in methanol ($K_{\mu}$ ranging from $-33$ to $-1$)~\cite{Jansen:PRL,Muller:AA,Ellingsen:ApJl}, at high redshift, have resulted in more stringent limits on the variation of $\mu$.

In this paper, we discuss the sensitivity of rotational transitions in CH and its deuterated isotopologue, CD to a variation of the proton-to-electron mass ratio, $\mu$, and the fine structure constant, $\alpha$. CH is a small diatomic radical that is frequently targeted in astrophysical studies, as it is a well established and well understood proxy of H$_2$~\cite{sheffer}. These studies have been targeting primarily the interstellar medium in the local galaxy. However, a survey for CH at high redshift is currently being conducted at the Atacama Large Millimeter Array (ALMA)~\cite{Muller:pc}. CH and CD have a spin-orbit coupling constant, $A$, that is close to two and four times their respective rotational constant, $B$. This leads to near degeneracies between rotational levels of different spin-orbit manifolds. As a result, the rotational transitions between the near-degenerate levels have an increased sensitivity to a variation of $\mu$. The work presented in this paper is complementary to that of Kozlov~\cite{Kozlov}, who calculated the sensitivity coefficients of Lambda-doublet transitions in CH and other diatomic radicals.

\section{Energy level structure of a $^{2}\Pi$ state}
\label{sec:simplified}

In this work, we investigate CH and CD in their $^{2}\Pi$ ground state. Molecules in $^{2}\Pi$ states have three angular momenta that need to be considered; the electronic orbital angular momentum, $\bf{L}$, the spin angular momentum, $\bf{S}$, and the rotational angular momentum, $\bf{R}$. Depending on the energy scales associated with these momenta, the coupling between the vectors is described by the different Hund's cases. In Hund's case (a), $\bf{L}$ is strongly coupled to the internuclear axis and $\bf{S}$ couples to $\bf{L}$ via spin-orbit interaction. States are labeled by $J$, the quantum number associated with the total angular momentum, and $\Omega$, the sum of $\Lambda$ and $\Sigma$, the projections of $\bf{L}$ and $\bf{S}$ on the internuclear axis, respectively. When the rotational energy becomes comparable to the energy of the spin-orbit interaction, $\bf{S}$ decouples from the internuclear axis, and Hund's case (b) is more appropriate. In this case the molecular levels are labelled by $N=R+\Lambda$, and $J$.

In heavy molecules at low $J$, the spin-orbit interaction is much larger than the rotational energy splitting. As a result, the energy level structure consists of two spin-orbit manifolds separated by an energy $A$, each having a pattern of rotational levels with energies given by $Bz$, with $z=(J+1/2)^2-1$. In light molecules, $A \sim Bz$ already at low $J$. In this case the two manifolds are considerably mixed and the energies are not described by a simple formula. In order to describe a situation that is intermediate between Hund's case (a) and (b), the wavefunction of a state is written as a superposition of pure Hund's case (a) wavefunctions:

\begin{equation}
|\Omega',J\rangle=
c_{\Omega',J,\Omega=1/2}|\Omega=1/2,J\rangle + 
c_{\Omega',J,\Omega=3/2}|\Omega=3/2,J\rangle 
\label{wavefunction}
\end{equation}

\noindent
where $c_{\Omega',J,\Omega=1/2}$ and $c_{\Omega',J,\Omega=3/2}$ are coefficients signifying the $\Omega=1/2$ and $\Omega=3/2$ character, respectively, of the wavefunction of the state $|\Omega',J\rangle$. Note that $\Omega'$ is used to label the rotational levels of the different spin-orbit manifolds, while $\Omega$ is used to denote the pure Hund's case (a) wavefunctions. The coefficients are found by diagonalizing the Hamiltonian matrix that is given, for instance, by Amiot \emph{et al.}~\cite{amiot}. When the Lambda-doublet splitting, centrifugal distortion and hyperfine splitting are neglected, the Hamiltonian matrix reduces to~\cite{brownevenson}:

\begin{equation}
 \left( \begin{array}{cc}
\frac{1}{2}A+Bz & -B\sqrt{z} \\
-B\sqrt{z} & -\frac{1}{2}A+B(z+2) \\ \end{array} \right) .
\label{H}
\end{equation}

\noindent 
The first row (or column) refers to the $^2\Pi_{\Omega'=3/2}$ component, the second to the $^2\Pi_{\Omega'=1/2}$ component. Although most of our calculations use the extensive matrix, all relevant features can be understood from the simplified matrix.

\begin{figure}
\includegraphics[width=\linewidth]{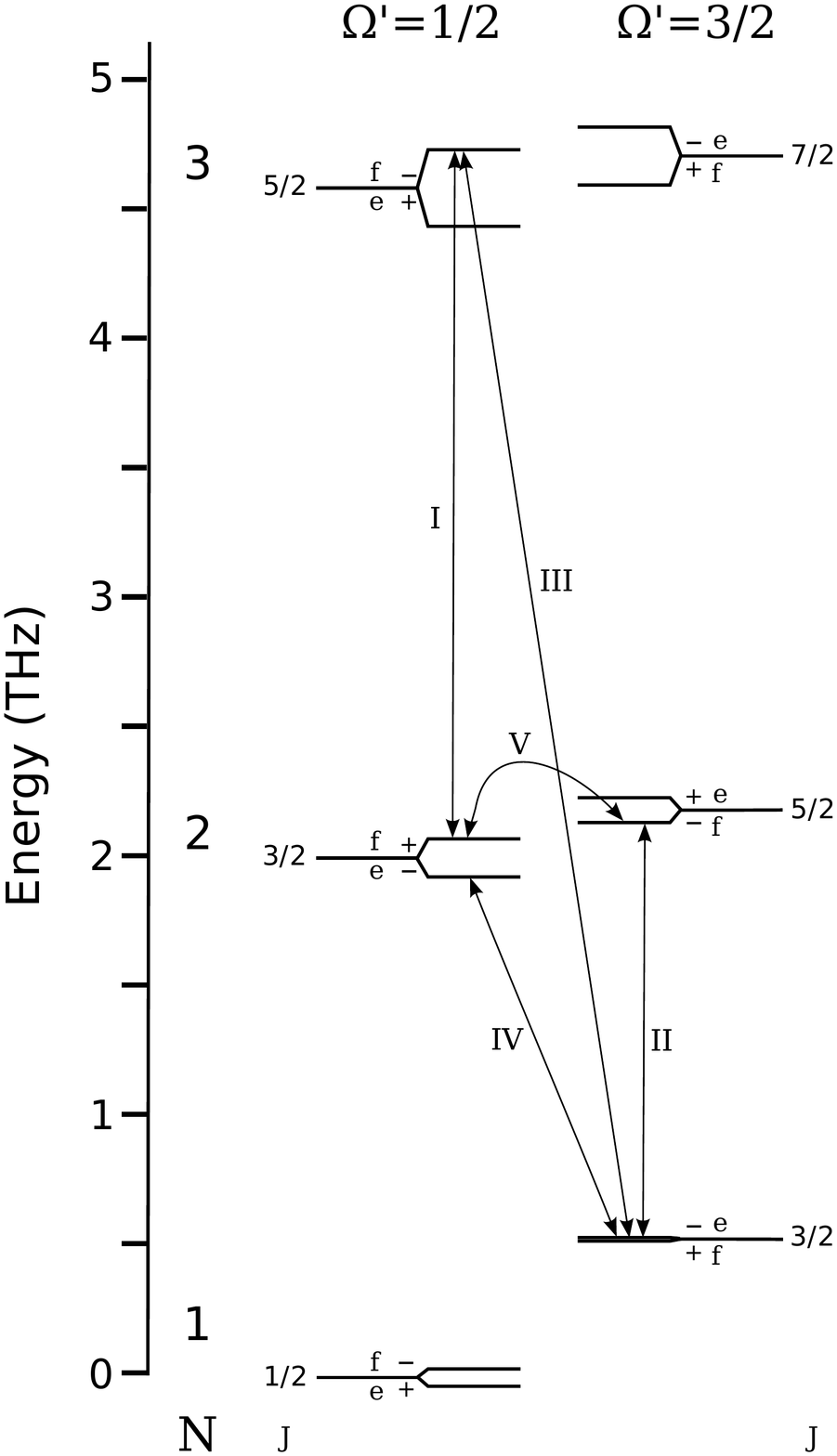}
\caption{Level scheme of the ground state of CH, calculated using the Hamiltonian matrix from Amiot \emph{et al.}~\cite{amiot} and the molecular constants given by McCarthy \emph{et al.}~\cite{mccarthy}. Indicated are five different types of rotational transitions, labeled \textsc{I} through \textsc{V}. The Lambda-doublet splitting is exaggerated by a factor ten. Also indicated are the symmetry of the electronic part of the wave function, denoted by \textit{e} and \textit{f} and the total parity, denoted by $+$ and $-$.}
\label{fig:levelscheme}
\end{figure}

The level scheme of CH is depicted in Fig.~\ref{fig:levelscheme}. In CH ($A=1.98B$), the $\Omega'=1/2,J=3/2$ level lies about 200~GHz below the $\Omega'=3/2,J=5/2$ level, whereas in CD ($A=3.65B$) the energy difference is only 30~GHz. In Fig.~\ref{fig:levelscheme}, the Lambda-doublet splittings are exaggerated by a factor of $10$. It was shown by Kozlov~\cite{Kozlov} that, as a result of an inversion of the Lambda-doublet splitting in the $\Omega'=3/2$-manifold, the different components of the Lambda doublet become near-degenerate at $\Omega'=3/2,J=3/2$ for CH, leading to enhanced sensitivity coefficients of the Lambda-doublet transitions.

Let us now consider the sensitivity of rotational transitions to a possible variation of $\mu$. The sensitivity coefficient of a transition is defined as

\begin{equation}
K_{\mu} = \frac{\mu}{\nu}\frac{\partial \nu}{\partial \mu}
= \frac{\mu_{\mathrm{\textit{red}}}}{\nu}
\frac{\partial \nu}{\partial \mu_{\mathrm{\textit{red}}}}
\label{K}
\end{equation}

\noindent
with

\begin{equation}
\nu = (E_{\Omega_{f}',J_{f}} - E_{\Omega_{i}',J_{i}})/h 
\label{nu}
\end{equation}

\noindent
the transition frequency, and $\mu_{\mathrm{\textit{red}}}$ the reduced mass of the molecule. Note that it is assumed here that the neutron and proton masses vary in the same way. The $K_{\mu}$ and $K_{\alpha}$ coefficients can now be calculated using the Hamiltonian matrix by including the dependence of the molecular constants on the reduced mass of the molecule and $\alpha$, given, for instance, in Beloy \textit{et al.}~\cite{Beloy:PRA}, and the values of the molecular parameters for CH from McCarthy \emph{et al.}~\cite{mccarthy} and for CD from Halfen \emph{et al.}~\cite{halfen}. As the effective Hamiltonian used for these molecules is an accurate physical representation, the sensitivity coefficients that are found these way are very accurate. For instance, in previous work on CO, the transition frequencies in different isotopologues could be predicted well within 10$^{-4}$\cite{denijs:pra2011}. However, for actual tests of the variation of fundamental constants, an accuracy of 1\% is well sufficient and the sensitivity coefficients will be given to this level only.

We have calculated $K_{\mu}$ and $K_{\alpha}$ for rotational transitions in CH and CD using both the extensive and the reduced matrix. For clarity, we separate the transitions into five different types, \textsc{I} through \textsc{V}, as shown in Fig.~\ref{fig:levelscheme}. Transitions from $J$ to $J+1$ within the $\Omega'=1/2$ and $\Omega'=3/2$ manifolds are labeled by \textsc{I} and \textsc{II}, respectively. Transitions from $\Omega'=1/2$ to $\Omega'=3/2$ with $\Delta J=-1,0,+1$ are labeled by \textsc{III}-\textsc{V}, respectively. From the calculations, we found that for both CH and CD, transitions of type~\textsc{I}-\textsc{II} have $K_{\mu}$ close to $-1$ and $K_{\alpha}$ close to $0$. Transitions of type~\textsc{III}-\textsc{IV} also have $K_{\mu}$ close to $-1$ and $K_{\alpha}$ close to $0$, except for transitions involving the lowest rotational levels, which have $K_{\mu}$ between $-0.5$ and $-1$ and a $K_{\alpha}$ between $1$ and $0$. Interestingly, transitions of type \textsc{V} were found to be extremely sensitive to variation of $\alpha$ and $\mu$. The $K_{\mu}$ for these transitions are listed in the third column of Table~\ref{tab:data} and range from $-67$ to $18$ for CD and $-6.2$ to $2.7$ for CH. The fourth column of Table~\ref{tab:data} lists the values of $K_{\alpha}$. Note, that $K_{\alpha}$ $\sim$ 2+2$K_{\mu}$, a relation that is exact when Lambda-type doubling is neglected.

\begin{figure}
\includegraphics[width=\linewidth]{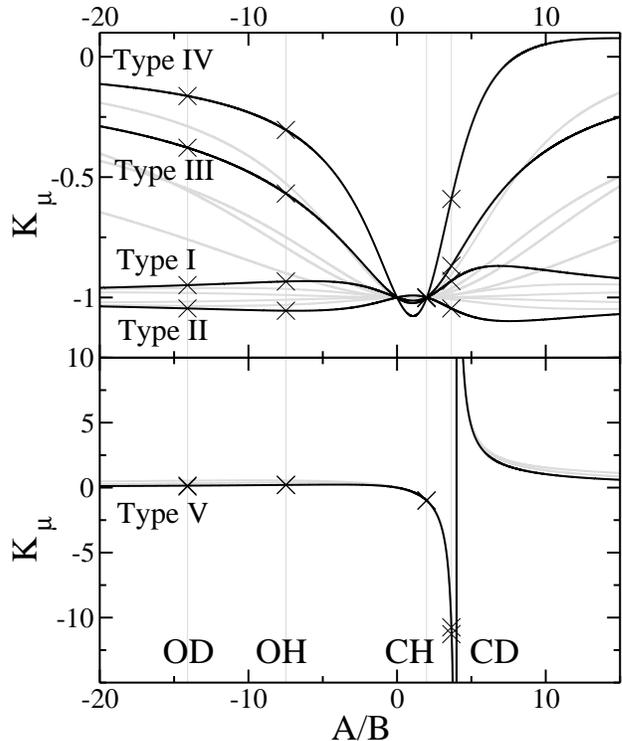}
\caption{Sensitivity coefficient $K_{\mu}$ of transition types \textsc{I}-\textsc{V} starting from $J=3/2$, in black, and $J=7/2$ and $J=15/2$, in gray, calculated using the reduced matrix given in Eq.~\ref{H}. The crosses indicate the sensitivity coefficients calculated for the transitions from $J=3/2$ for the listed molecules using the complete matrix.}
\label{fig:sens}
\end{figure}

The calculations are most easily understood by plotting the sensitivity coefficients for the different transitions as a function of $A/B$, as shown in Fig.~\ref{fig:sens}. The upper panel shows the $K_{\mu}$ for transitions of types~\textsc{I}-\textsc{IV}, while the lower panel shows $K_{\mu}$ for transitions of type \textsc{V}, calculated using the reduced matrix from Eq.~\ref{H}. The black curves show the sensitivity coefficients for transitions starting from $J=3/2$. To indicate the progression towards higher values of $J$, transitions starting from $J=7/2$ and $J=15/2$ are plotted in gray. We see that for large $|A/B|$, $K_{\mu}$ approaches $-1$ for transitions of type \textsc{I} and \textsc{II} and $0$ for transitions of type \textsc{III}-\textsc{V}. This can be understood by realizing that for large $|A/B|$, a Hund's case (a) coupling scheme applies. Consequently, transitions of type~\textsc{I} and \textsc{II} are pure rotational transitions which are proportional to $B$, while transitions of type \textsc{III}-\textsc{IV} are pure electronic transitions and proportional to $A$. When $A \sim Bz$, the manifolds become mixed and the sensitivity of the different types of transition is between $0$ and $-1$. When $A=0$, corresponding to a pure Hund's case (b), all types of transitions have a sensitivity coefficient $K_\mu$ of $-1$, as expected. When $A=2B$ the two spin-orbit manifolds are fully mixed, also causing $K_\mu$ to become $-1$. Another special case is when $A=4B$. Here, $\Omega'=3/2, J$ levels are degenerate with $\Omega'=1/2, J+1$ levels. This gives rise to an enhancement of the sensitivity coefficient for transitions that connect these levels, i.e., transitions of type \textsc{V}. The enhancement is expected to be on the order of $A/\nu$~\cite{Bethlem:FarDisc,Jansen:PRA}, which is in reasonable agreement with our calculations. Note, that the sensitivity coefficients found using the simplified model are almost independent of $J$.

The crosses also shown in Fig.~\ref{fig:sens} are the values of $K_{\mu}$  calculated using a full set of molecular parameters for CH ($A=1.98B$), CD ($A=3.65B$), OH ($A=-7.48B$)~\cite{brownradford} and OD ($A=-14.1B$)~\cite{brownschubert}. The correspondence between the simplified model and the full description is very good for transitions at low $J$, but less good for higher $J$ when effects of the Lambda-type doubling become increasingly important. The Lambda-type doubling shifts the energy levels, leading to a decrease or increase of the energy difference between the $\Omega'=1/2, J$ and $\Omega'=3/2, J+1$ levels, and henceforth to a corresponding increase or decrease of the sensitivity coefficients.

\section{Transition strengths}
\label{sec:transstr}

In order to be relevant for astrophysical tests of the time-variation of the proton-to-electron mass ratio, the highly sensitive transitions in CH and CD discussed in the previous section need to be sufficiently strong. In Hund's case (a), transitions between different $\Omega$ manifolds, i.e., transitions of types \textsc{III}-\textsc{V}, are forbidden. However, as discussed in the previous section, the $\Omega$ manifolds of CH and CD are mixed and transitions are allowed.

The transition strength of a transition between rotational states $i$ and $f$, is given by $|\langle i|T|f\rangle|^2$, with $|i\rangle$ and $|f\rangle$ given by Eq.~(\ref{wavefunction}). The transition strength of a transition $i \rightarrow f$ is then given by:

\begin{equation}
\begin{split}
|\langle i|T|f \rangle|^2 = | & c_{i,1/2} \cdot c_{f,1/2}\langle 1/2,J_i |T| 1/2,J_f \rangle+\\
& c_{i,3/2} \cdot c_{f,1/2}\langle 3/2,J_i |T| 1/2,J_f \rangle+\\
& c_{i,1/2} \cdot c_{f,3/2}\langle 1/2,J_i |T| 3/2,J_f \rangle+\\
& c_{i,3/2} \cdot c_{f,3/2}\langle 3/2,J_i |T| 3/2,J_f \rangle|^2.
\end{split}
\label{eq:ab}
\end{equation}

\begin{figure}
\includegraphics[width=\linewidth]{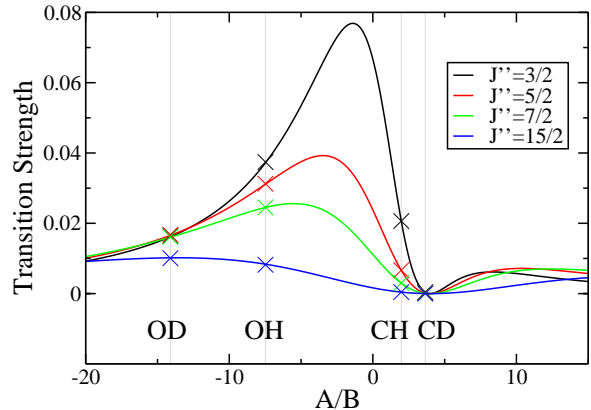}
\caption{Transitions strengths of type \textsc{V} transitions, following from Eq.~\ref{eq:ab}, starting from different $J$ levels. The transition strength is zero at $A=4B$ for all $J$, making the transitions with the highest sensitivity exceedingly weak. The crosses correspond to a calculation for CH, CD, OH and OD using a complete set of parameters. The molecules are also indicated by the vertical grey lines.}
\label{fig:transstrength}
\end{figure}

\noindent
The expressions $\langle \Omega,J_i |T| \Omega,J_f \rangle$ are the Hund's case (a) dipole transition matrix elements given in, for example, Brown and Carrington \cite{browncarrington}. As a result of the Hund's case (a) selection rules, the second and third terms on the right-hand side of Eq.~(\ref{eq:ab}) are zero. Using the simplified Hamiltonian matrix given in Eq.~(\ref{H}), we have calculated the amplitude of the remaining terms as a function of $A/B$. In Fig.~\ref{fig:transstrength}, the transition strength is plotted for type \textsc{V} transitions starting from different $J$ levels. It is seen that when $|A/B|$ becomes smaller, the levels become increasingly mixed and the transition strength becomes larger. Near $A=4B$ the transition strength becomes smaller, due to destructive interference between the two different paths that combine to form this transition. At $A=4B$ the two paths are equally strong, but due to orthogonality of the eigenvectors they have a different sign and the transition strength becomes zero. The last column of Table~\ref{tab:data} lists the transition strength, calculated using the full Hamiltonian, but neglecting hyperfine splitting. For comparison, note that purely rotational transitions have a transition strength of order unity. The crosses shown in Fig.~\ref{fig:transstrength} again correspond to a calculation for CH, CD, OH and OD using a complete set of parameters and are in good agreement with the calculations using the reduced matrix. We have validated that these calculations are also in agreement with calculations using the PGopher software package~\cite{pgopher}. Note, that the dipole moment is set to unity in the calculations.

\begin{table}
\begin{ruledtabular}
\caption{Transition frequencies, sensitivity coefficients to variation of $\mu$ and $\alpha$ and transition strengths of transitions from $\Omega'=1/2, J$ to $\Omega'=3/2, J+1$, type \textsc{V} transitions in CH and CD calculated using the Hamiltonian matrix from Amiot \emph{et al.}~\cite{amiot} and the molecular constants given by McCarthy \emph{et al.}~\cite{mccarthy} for CH and Halfen \emph{et al.}~\cite{halfen} for CD. Note that the values of $K_{\mu}$ for transitions starting from $\Omega'=1/2, J=1/2$ is always between $0$ and $-1$, as the $\Omega'=1/2, J=1/2$ is unmixed. freq.: Frequency, Tr. Str.: Transition Strength.}
\begin{tabular}{lldlcc}
&$J$	&\mathrm{freq}_\cdot \mathrm{(MHz)}&	$K_\mu$&	$K_\alpha$	&Tr. Str. \\
\hline
\bigstrut
CH &$e$-parity&&&&\\
&1/2   &  536772.4   &  -0.22  &1.57 &   6.6$\cdot10^{-1}$      \\
&3/2   &  191101.3 &  -1.02 & -0.0068 &  2.1$\cdot10^{-2}$  \\
&5/2   &  137163.5 &  -1.09 & -0.041 &   6.7$\cdot10^{-3}$  \\
&7/2   &  115440.4 &  -1.20 & -0.074 &   3.0$\cdot10^{-3}$  \\
&9/2   &  107620.7 &  -1.32 & -0.10 &    1.6$\cdot10^{-3}$  \\
&11/2  &  107870.7 &  -1.44 & -0.12 &    9.7$\cdot10^{-4}$  \\
&13/2  &  113649.5 &  -1.55 & -0.14 &    6.3$\cdot10^{-4}$  \\
&15/2  &  123632.5 &  -1.65 & -0.14 &    4.3$\cdot10^{-4}$  \\
\hline
\bigstrut
&      $f$-parity&&&&\\
&1/2   &    532741.0   &  -0.20  & 1.59 &    6.6$\cdot10^{-1}$      \\
&3/2   &    178904.5  &  -0.94  &0.039 &     2.1$\cdot10^{-2}$  \\
&5/2   &    111119.2  &  -0.85  &0.020 &     6.7$\cdot10^{-3}$  \\
&7/2   &    71064.4  &  -0.64  &-0.012 &     3.0$\cdot10^{-3}$  \\
&9/2   &    40500.9  &  -0.11  &-0.086 &     1.6$\cdot10^{-3}$  \\
&11/2  &    13698.2  &   2.56  &-0.42 &     9.7$\cdot10^{-4}$  \\
&13/2  &    11758.8  &  -6.40  &0.68 &     6.3$\cdot10^{-4}$  \\
&15/2  &    37049.6  &  -3.16  &0.27 &     4.3$\cdot10^{-4}$  \\
 \hline
 \bigstrut
CD   &  $e$-parity  &&&& \\
&1/2  &  439799.0  &  -0.46  & 1.09 & 9.5$\cdot10^{-1}$      \\
&3/2  &  31493.8  &  -10.6  &-19.1  &    4.1$\cdot10^{-4}$  \\
&5/2  &  23326.3  &  -10.0  &-17.9  &   1.7$\cdot10^{-4}$  \\
&7/2  &  20438.4  &  -9.09  &-15.8  &   8.6$\cdot10^{-5}$  \\
&9/2  &  20133.3  &  -7.91  &-13.2  &   4.9$\cdot10^{-5}$  \\
&11/2  &  21473.9  &  -6.74  &-10.6  & 3.1$\cdot10^{-5}$  \\
&13/2  &  24037.3  &  -5.71  &-8.27  & 2.1$\cdot10^{-5}$  \\
&15/2  &  27598.2  &  -4.89  &-6.44  & 1.5$\cdot10^{-5}$  \\
 \hline
 \bigstrut
&$f$-parity&&&&\\
&1/2  &  439262.1  &  -0.45  &  1.10&9.5$\cdot10^{-1}$      \\
&3/2  &  29320.6  &  -11.1  &-20.3  &4.2$\cdot10^{-4}$  \\
&5/2  &  17073.0  &  -12.7  &-24.0  &1.7$\cdot10^{-4}$  \\
&7/2  &  8787.4  &  -18.0  &-35.9  &  8.6$\cdot10^{-5}$  \\
&9/2  &  1771.6  &  -67.1  &-146  &4.9$\cdot10^{-5}$  \\
&11/2 &  4894.4  &  18.0  &44.7  &      3.1$\cdot10^{-5}$  \\
&13/2 &  11611.1  &  5.38  &16.4  &    2.1$\cdot10^{-5}$  \\
&15/2 &  18576.6  &  2.12  &9.14  &    1.5$\cdot10^{-5}$  \\
\end{tabular}
\label{tab:data}
\end{ruledtabular}
\end{table}

\section{Relevance for tests on drifting constants over cosmological time scales}
\label{sec:relevance}

In the previous section, we have shown that transitions of type $\textsc{V}$ that have an enhanced sensitivity to a variation of $\mu$ are too weak to be observed in astrophysical objects at high redshift. The only transitions in CH that have  a non-vanishing transition strength and a $K_{\mu}$ that deviates significantly from $-1$ are the $\Omega'=1/2,J=1/2$ to $\Omega'=3/2,J=3/2$ transitions at $532$ and $536$~GHz that have $K_{\mu}=-0.2$. By comparing these transitions with a rotational transition, typically exhibiting $K_{\mu}=-1$, in any other molecule observed in the same object, a test of the time-variation of $\mu$ over cosmological time scales can be performed. If $\mu$ varies, the transition frequency of a pure rotational transition will vary while the frequency of the discussed transition in CH will change five times less; i.e., the CH transition will act as an anchor line. Ideally, the CH anchor transitions are compared with other transitions in CH, and preferably with transitions from the same levels. 
This eliminates one of the main systematic effects that limits astrophysical tests, namely the effect of spatial segregation. Astrophysical tests rely on the assumption that the transitions that are being compared originate from the same location and hence the same apparent redshift. Spatial segregation of the absorbers may mimic or hide frequency shifts due to a variation of $\mu$~\cite{Murphy:Science}.

We propose a test of the time-variation of $\mu$ by comparing the CH anchor transitions to other rotational or Lambda-doublet transitions in CH involving the same absorbing ground levels, i.e. to the $\Omega'=1/2,J=1/2$ to $\Omega'=1/2,J=3/2$ transition near 2~THz and/or the $\Omega'=3/2,J=3/2$ to $\Omega'=3/2,J=5/2$ transition near 1.5~THz that have $K_{\mu}=-1$ or to the Lambda-doublet transition in the $\Omega'=1/2,J=1/2$ at 3.3~GHz that has $K_{\mu}=-1.7$ and the Lambda-doublet transition in the $\Omega'=3/2,J=3/2$ near 700~MHz that has $K_{\mu}=-6.2$~\cite{Kozlov}. This test is based on transitions within the lowest four levels of a single species making it very robust against possible shifts due to spatial segregation of the absorbing molecules. The transitions that are relevant to this test are listed in Table~\ref{tab:datahyp}, including the hyperfine splitting, with their respective sensitivity coefficients and transitions strengths, calculated using PGopher~\cite{pgopher}. Our values for the sensitivity coefficients of the Lambda-doubling transitions correspond well to those found by Kozlov~\cite{Kozlov}, but our sensitivity coefficients are more exact as we use a more complete set of molecular parameters.

\begingroup
\squeezetable
\begin{table}
\begin{ruledtabular}
\caption{Transition frequencies, sensitivity to variation of $\mu$ and $\alpha$ and transitions strengths of specific Lambda doubling and rotational transitions in CH calculated using PGopher~\cite{pgopher} with the molecular constants from McCarthy \emph{et al.}~\cite{mccarthy}, including hyperfine splitting. Measured frequencies are given where possible, the difference with calculations is given for these transitions. The letters correspond to references: a: McCarthy \emph{et al.}~\cite{mccarthy}, b: Brazier and Brown~\cite{brazier}, c: Ziurys and Turner~\cite{ziurys}, d: Amano~\cite{amano}, e: Davidson \emph{et al.}~\cite{davidson}. freq.: Frequency, o-c: Observed - Calculated, Trans. Str.: Transition Strength.}
\begin{tabular}{lldrlll}
$\Omega'$, J	&F&\mathrm{freq}_\cdot&o-c~~~~&	$K_\mu$&	$K_\alpha$&Trans.\\
&&\mathrm{(MHz)}&(kHz)&&&Str.\\
\hline
Lambda-doubling&&&&&&\\
\hline
\bigstrut
$\frac{1}{2}, \frac{1}{2}$ $f$ $\rightarrow$ $\frac{1}{2}, \frac{1}{2}$ $e$  &  0$\rightarrow$1	&	3263.795^a	&16~~&	-1.71	&	0.58	&	0.33	\\
          &  1$\rightarrow$1	&	3335.481^a	&-10~~&	-1.70	&0.61&0.67	\\
          &  1$\rightarrow$0	&	3349.194^a	&6~~&	-1.69	&	0.62	&0.33	\\
$\frac{1}{2}, \frac{3}{2}$ $f$ $\rightarrow$ $\frac{1}{2}, \frac{3}{2}$ $e$  &  1$\rightarrow$2	&	7275.004^a	&15~~&	-2.13	&	-0.26	&	0.14	\\
              &  1$\rightarrow$1	&	7325.203^a &27~~&	-2.12	&	-0.24	&0.68	\\
              &  2$\rightarrow$2	&	7348.419^a	&-15~~&	-2.12	&	-0.24	&1.23	\\
              &  2$\rightarrow$1	&	7398.618^a	&-4~~&	-2.11	&	-0.22	&0.14	\\
$\frac{1}{2}, \frac{5}{2}$ $f$ $\rightarrow$ $\frac{1}{2}, \frac{5}{2}$ $e$  &  2$\rightarrow$3	&	14713.78^b	& 190~~&	-2.02	&	-0.04	&	0.04	\\
              &  2$\rightarrow$2	&	14756.670^a	&36~~&	-2.01	&	-0.03	&0.54	\\
              &  3$\rightarrow$3	&	14778.962^a	&-28~~&	-2.01	&	-0.03	&0.77	\\
              &  3$\rightarrow$2	&	14821.88^b	&-160~~&	-2.01	&	-0.02	&0.04	\\
$\frac{3}{2}, \frac{3}{2}$ $f$ $\rightarrow$ $\frac{3}{2}, \frac{3}{2}$ $e$  &  2$\rightarrow$2	&	701.667^c		      &-8~~&	-6.14	&	-8.28	&	1.17	\\
              &  1$\rightarrow$2	&	704.008		&--~~&	-6.11	&-8.23&0.13	\\
              &  2$\rightarrow$1	&	722.452  &--~~&	-5.98	&-7.96&0.13	\\
              &  1$\rightarrow$1	&	724.788^c  &3~~&	-5.96	&-7.92&0.65	\\
Rotational&&&&&&\\
\hline
\bigstrut
$\frac{1}{2}, \frac{1}{2}$ $f$ $\rightarrow$ $\frac{3}{2}, \frac{3}{2}$ $f$  &  1$\rightarrow$1	&	532721.333^d	&-314~~&	-0.20	&	1.59	&	0.17	\\
                              &  1$\rightarrow$2	&	532723.926^d	&-54~~&	-0.20	&1.59&0.85	\\
                              &  0$\rightarrow$1	&	532793.309^d	&-50~~&	-0.20	&1.59&0.34	\\
$\frac{1}{2}, \frac{1}{2}$ $e$ $\rightarrow$ $\frac{3}{2}, \frac{3}{2}$ $e$  &  1$\rightarrow$2	&	536761.145^d	&-1~~&	-0.22	&	1.57	&	0.86	\\
                              &  1$\rightarrow$1	&	536781.954^d	&31~~&	-0.22	&1.57&0.17	\\
                              &  0$\rightarrow$1	&	536795.678^d	&58~~&	-0.22	&1.57&0.34	\\
$\frac{3}{2}, \frac{3}{2}$ $f$ $\rightarrow$ $\frac{1}{2}, \frac{3}{2}$ $e$  &  2$\rightarrow$1	&	1470689.444	&--~~&	-1.00	&	 0.00	&	0.03	\\
                              &  1$\rightarrow$1	&	1470691.777	&--~~&	-1.00	&0.00&0.17	\\
                              &  2$\rightarrow$2	&	1470739.632	&--~~&	-1.00	&0.00&0.30	\\
                              &  1$\rightarrow$2	&	1470741.965	&--~~&	-1.00	&0.00&0.03	\\
$\frac{3}{2}, \frac{3}{2}$ $e$ $\rightarrow$ $\frac{1}{2}, \frac{3}{2}$ $f$  &  1$\rightarrow$1	&	1477292.168	&--~~&	-1.00	&	0.00	&	0.16	\\
                              &  2$\rightarrow$1	&	1477312.946	&--~~&	-1.00	&0.00&0.03	\\
                              &  1$\rightarrow$2	&	1477365.614	&--~~&	-1.00	&0.00&0.03	\\
                              &  2$\rightarrow$1	&	1477386.391	&--~~&	-1.00	&0.00&0.30	\\
$\frac{3}{2}, \frac{3}{2}$ $f$ $\rightarrow$ $\frac{3}{2}, \frac{5}{2}$ $f$  &  2$\rightarrow$3	&	1656961.185	&--~~&	-1.00	&	0.00	&	2.32	\\
                              &  2$\rightarrow$2	&	1656970.448	&--~~&	-1.00	&0.00&0.17	\\
        			&  1$\rightarrow$2	&	1656972.781	&--~~&	-1.00	&0.00&1.49	\\
$\frac{3}{2}, \frac{3}{2}$ $e$ $\rightarrow$ $\frac{3}{2}, \frac{5}{2}$ $e$   &  2$\rightarrow$3	&	1661107.278	&--~~&	-1.00	&	0.00	&	2.32	\\
				&  1$\rightarrow$2	&	1661118.045	&--~~&	-1.00	&0.00&1.49	\\
				&  2$\rightarrow$2	&	1661138.822	&--~~&	-1.00	&0.00&0.17	\\
$\frac{1}{2}, \frac{1}{2}$ $e$ $\rightarrow$ $\frac{1}{2}, \frac{3}{2}$ $e$  &  1$\rightarrow$1	&	2006748.915	&--~~&	-0.79	&	0.42	&	0.16	\\
                              &  0$\rightarrow$1	&	2006762.612	&--~~&	-0.79	&0.42&0.32	\\
                              &  1$\rightarrow$2	&	2006799.103	&--~~&	-0.79	&0.42&0.81	\\
$\frac{1}{2}, \frac{1}{2}$ $f$ $\rightarrow$ $\frac{1}{2}, \frac{3}{2}$ $f$  &  1$\rightarrow$1	&	2010738.601	&--~~&	-0.79	&	0.42	&	0.16	\\
                              &  0$\rightarrow$1	&	2010810.46^e	&150~~&	-0.79	&0.42&0.33	\\
                              &  1$\rightarrow$2	&	2010811.92^e	&-130~~&	-0.79&0.42&0.81	\\
\end{tabular}
\label{tab:datahyp}
\end{ruledtabular}
\end{table}
\endgroup

\section{Conclusion}
\label{sec:conclusion}

In this paper, we have analyzed the sensitivity to a possible variation of $\mu$ and $\alpha$ for rotational transitions in $^2 \Pi$ states, in particular for rotational transitions in the groundstate of CH and CD. For certain rotational transitions, we found a significantly enhanced sensitivity due to accidental degeneracies between rotational levels of different fine-structure manifolds. These degeneracies occur when the spin-orbit coupling constant is close to four times the rotational constant. CH ($A=1.98B$) and particularly CD ($A=3.65B$), match this condition closely. The fact that enhancement occurs is unexpected, as it was shown by Bethlem and Ubachs~\cite{Bethlem:FarDisc} that in molecules such as CO, the transition from Hund's (a) to Hund's case (b) coupling scheme prohibits levels that are connected by one-photon transitions to be become near-degenerate. Here we show that for $A \sim 4B$ this does not apply. Unfortunately, the same condition that leads to an enhanced sensitivity suppresses the transition strength. Thus, one-photon transitions between different spin-orbit manifolds of molecular radicals are either insensitive or too weak to be of relevance for tests of the variation of fundamental constants over cosmological time scales. However, the high sensitivity coefficients could possibly be used in laboratory tests (note that experiments are being planned to decelerate CH molecules using electric fields~\cite{Tarbutt:pc} which open the prospect of measuring its rotational and microwave spectrum at high resolution).

We propose a test in CH based on the comparison between the rotational transitions between the $e$ and $f$ components of the $\Omega'=1/2,J=1/2$ and $\Omega'=3/2,J=3/2$ levels at $532$ and $536$\,GHz which have $K_\mu=-0.2$, with other rotational or Lambda-doublet transitions in CH. Such a test, to be performed by far infrared spectroscopy of highly redshifted objects, is robust against systematic effects.

\section{Acknowledgements}
We thank Sebastien Muller for helpful discussions. This work is financially supported by the Netherlands Foundation for Fundamental Research of Matter (FOM) (project 10PR2793 and program ``Broken mirrors and drifting constants''). H.L.B. acknowledges financial support from NWO via a VIDI-grant, and from the ERC via a Starting Grant.

\bibliography{bib}

\end{document}